\documentclass[11pt]{article}\hfuzz=10pt  \oddsidemargin=0.2cm \textheight 223mm \textwidth=16.2cm
\newcommand{\comm}[1]{}

\def\xxxonly{\rd}
\def\xxxonly{}\def\xxxonly{\comm}

\def\noxxx{\comm}

\usepackage{amssymb}
\usepackage{amsmath}\usepackage{graphicx}
\usepackage{times}\hfuzz=10pt 
 \usepackage{epsfig}
  \usepackage{color}

\usepackage[numbers]{natbib}
\def\citet{\cite}
 
\newtheorem{theorem}{Theorem}
\newtheorem{lemma}{Lemma}
\newtheorem{proposition}{Proposition}
\newtheorem{definition}{Definition}
\newtheorem{remark}{Remark}
\newtheorem{example}{Example}

\def\e{\varepsilon}

\def\defi{\stackrel{{\scriptscriptstyle \Delta}}{=}}

\def\a{\alpha}
\def\d{\delta}
\def\o{\omega}
\def\O{\Omega}

\def\Y{{\cal Y}}
\def\F{{\cal F}}
\def\w{\widehat}
\def\Ind{{\mathbb{I}}}

\def\mes{{\rm mes\,}}

\def\esssup{\mathop{\rm ess\, sup}}

\def\Re{{\rm Re\,}}

\def\R{{\bf R}}

\def\F{{\cal Z}}

\def\L{L}
\def\F{{\Feta}}

\def\g{\gamma}
\def\C{{\bf C}}

\def\ww{\widetilde}
\def\X{{\cal X}}

\def\oo{\bar}

\def\G{\Gamma}

\def\V{{\cal V}}

\def\L{{\cal L}}

\def\T{{\mathbb{T}}}

\def\k{\kappa}

\newcommand{\be}{\begin{equation}}
\newcommand{\ee}{\end{equation}}
\newcommand{\bd}{\begin{displaymath}}
\newcommand{\ed}{\end{displaymath}}
\newcommand{\ba}{\begin{array}{ll}}
\newcommand{\ea}{\end{array}}
\newcommand{\baa}{\begin{eqnarray}}
\newcommand{\eaa}{\end{eqnarray}}
\newcommand{\baaa}{\begin{eqnarray*}}
\newcommand{\eaaa}{\end{eqnarray*}}

\def\oo{\bar}

\def\a{\alpha}

\def\K{{\cal K}}
\def\Ko{K}
\def\ko{\kappa}
\def\Yo{Y}
\def\yo{y}
\def\ZZ{{\bf Z}}
\def\F{{\cal F}}
\def\ew{\left(e^{i\o}\right)}
\def\ew{\left(i\o\right)}
\def\NN{\eta}
\def\circ{*}
\date{Submitted: May 8,  2017.  Revised: January 9, 2020}
\title{On  linear weak predictability with single point spectrum degeneracy}
\author{
Nikolai Dokuchaev}
 
\begin{document}
\def\breakkk{}%
\def\brea{}
\def\breakk{}
\maketitle
\noxxx{\let\thefootnote\relax\footnote{The author is with the ITMO University, St. Petersburg, 197101 Russian Federation the School of Electrical Engineering, Computing and Mathematical Sciences, Curtin
University,  GPO Box U1987, Perth, 6845 Western Australia.}}
\begin{abstract} The paper studies  properties of continuous time processes
with spectrum degeneracy at a single point where their Fourier transforms vanish with a certain rate. It appears that these processes are linearly predictable in some weak sense, meaning that   convolution integrals over future times can be approximated by causal convolutions over past times.  The corresponding predicting kernels are time invariant, and
they are  presented explicitly in the frequency domain via their transfer functions. These predictors  are "universal" meaning that they do not require to know details of the spectrum of the underlying processes; the same predictor can be used for the entire class of  processes with a single point spectrum degeneracy.   The predictors feature some robustness with respect to noise contamination.
\comm{\par MSC 2010 classification : 42A38, 
93E10, 
62M15,      
42B30. }  
\end{abstract}
\def\break{}
\def\break{\\ }
\section{Introduction} The paper studies properties of continuous time processes
with spectrum degeneracy in a pathwise deterministic setting, i.e.,  without probabilistic assumptions on the ensemble, where an underlying process is  deemed to be unique and such that one cannot rely on statistics  collected from observations of other similar paths.
A decision (a prediction, an estimate,  etc.) has to be based  on the intrinsic properties of this single  observed path.

There are some  opportunities for prediction and interpolation
of continuous time processes in pathwise setting with  certain degeneracy of
their spectrum.
\begin{itemize}
\item
In the stochastic setting for  continuous time stationary Gaussian  processes,
there exist optimal predictors represented by  causal linear integral operators; see the review
of these results in \cite{Dym,Yaglom}. The predictors are optimal in the sense of minimization of the mean square error;
their selection is defined by the  spectral density $\phi$ of the underlying process.  By the Kolmogorov-Krein Theorem, this error can be zero
 if and only if  \baa
\int_{-\infty}^\infty \frac{\log\phi(\o)}{1+\o^2} d\o=-\infty;
\label{KK}
\eaa
see, e.g., \cite{Ib}, p. 57.

\item The classical
Nyquist-Shannon-Kotelnikov interpolation theorem states  that
 a band-limited function can be uniquely recovered without error  from  a infinite equidistant
sampling sequence.  The sampling rate must be at least twice the maximum frequency
present in the signal (the critical Nyquist rate).

\item Functions  are uniquely defined by the  samples taken with the rate defined  by the measure
of the spectrum support only; see  \cite{La67}, p.39.

\item Functions  with certain periodicity of the location of gaps in the spectrum and with some restrictions
on the measure or on accumulation at infinity  of the spectrum gap
are uniquely defined by the  sparse samples below the Nyquist rate at sampling points  deviating slightly from arithmetic
progressions  \cite{La,La2,OU08,OU,U1,U2}.
\item Band-limited functions are analytic and are uniquely defined by  their values on an arbitrarily small time interval.
In particular, band-limited functions are  uniquely defined by their past values, i.e. predictable.
\item Functions with exponential decrease of energy on higher frequencies are uniquely defined by their past values. Moreover, there exist
linear predictors that do not require to know the spectrum, with the prediction horizon defined by the rate of the energy decrease  \citet{D10}.
\item Functions with the Fourier transform vanishing on an arbitrarily small interval $(-\O,\O)$ for some $\O>0$  are uniquely defined by their past values. There are
linear predictors defined by $\O$ only that  allow to predict anticausal convolutions involving the future values  \citet{D08}.
\end{itemize}

The present  paper shows that  a  degeneracy  of the Fourier transform for continuous processes  at a single point only still ensures some linear extrapolation opportunities for continuous time processes in the pathwise deterministic setting.
It shows  that processes featuring this degeneracy
are linearly predictable is some weak sense, meaning that anti-causal  convolution
integrals over future time can be approximated by causal convolution integrals over past time (Theorem \ref{Th1}).
 This result  sheds some  new light on the impact of
spectrum degeneracy on the predictability and extrapolation.

To prove the predictability of the anti-causal convolutions, we obtained a family of new linear predictors represented by
causal convolutions  (Theorem \ref{ThM}). The  predictors are given explicitly in the frequency domain.

The predictors suggested in the paper are not error free; however, the prediction error can be made arbitrarily
small, and there is some robustness   with respect to the noise contamination.
The predictors  suggested here are constructed using the approach developed in \cite{D08,D10,D12,D12b} but
are quite different.

We emphasize that this result is not a straightforward rewording linear of extrapolation results known for
stochastic Gaussian processes  with  the spectral densities. One reason for this is that the properties of these stationary processes are  quite special and cannot be
mechanically transferred to deterministic functions and their spectrums.  For example, it appears that  the criterion of recoverability of a single value  for a discrete time stationary Gaussian process is different than in the
pathwise deterministic setting (\cite{D17}, p.86).  Furthermore, the optimal extrapolating operators  known for Gaussian stationary processes have to be  constructed  for  a particular shape of the spectral density    (see e.g. \cite{Dym,W,M,V,Ly,Yaglom}).  On the other hand, unlike the linear predictors known for the Gaussian processes, the predictors introduced below are "universal" meaning that they do not require to know the shape of the spectrum (i.e. the Fourier transform) of the underlying processes; the same predictor can be used for a large class of different  processes.

The paper is organized in the following manner. In Section
\ref{secDef}, we formulate the definitions and background facts related to
the linear weak predictability. In Section
\ref{secM}, we formulate
 the main theorems on predictability and predictors (Theorem \ref{Th1} and Theorem \ref{ThM}).
Section \ref{secProof} contains the proofs. In Section \ref{secRob}, we discuss the
robustness of the predictors.
Finally, in Section \ref{secCon}, we discuss  our results.

\section{Definitions and background}\label{secDef}
Let  $\Ind$ denote the indicator function, $\R^+\defi[0,+\infty)$,
$\C^+\defi\{z\in\C:\ \Re z>  0\}$, $i=\sqrt{-1}$.
\par
For complex valued functions $x\in L_1(\R)$ or $x\in L_2(\R)$, we
denote by $\F x$ the function defined on $i\R$ as the Fourier
transform of $x$:
 $$(\F x)(i\o)= \int_{-\infty}^{\infty}e^{-i\o
t}x(t)dt,\quad \o\in\R.$$ If $x\in L_2(\R)$, then $X$ is defined
as an element of $L_2(\R)$ (meaning $L_2(i\R)$).

\par
For $x\in L_2(\R)$ such that $x(t)=0$ for $t<0$, we
denote by $\L x$  the Laplace transform \baa\label{Up} (\L
x)(z)=\int_{0}^{\infty}e^{-z t}x(t)dt, \quad z\in\C^+. \eaa

\par
Let  $H^p$ be the Hardy space of holomorphic on $\C^+$ functions
$Q(z)$ with finite norm
$\|Q\|_{H^p}=\sup_{s>0}\|Q(s+i\o)\|_{L_p(\R)}$, $p\in[1,+\infty]$;
see, e.g.,  \cite{Du}, Chapter 11.

By the Paley-Wiener Theorem,
$X\in H^2$  if and only if $X=\L x$ for some
 $x\in L_2(\R)$ such that $x(t)=0$ for $t<0$;
see e.g. Theorem 19.2 in \cite{Rudin}, p.372.

The definitions below  in this section are similar to the definitions introduced in \cite{D08}.
\begin{definition}
Let $\K$ be the class of functions $\kappa:\R\to\R$ such that $\kappa(t)=0$
for $t>0$ and such that, for any $\kappa\in\K$, there exists an integer $m>0$, a set $\{a_k\}_{k=1}^m\subset(0,+\infty)$, and a polynomial $d$ such that ${\rm deg\,} d < m$
and  $K=\F k$ is represented as \be
K(i\o)=\frac{d(i\o)}{\d(i\o)},\quad \o\in\R, \label{kda}\ee where $\d(i\o)\defi \prod_{j=1}^m (i\o-a_j)$.
\end{definition}
\par
In particular, the class $\K$ includes all
 linear combinations of functions $e^{\lambda t}\Ind_{\{t\le 0\}}$,
where $\lambda\in (0,+\infty)$.
\begin{definition}
Let $\w\K$ be the class of functions $\w\kappa:\R\to\R$ such that $\w k
(t)=0$ for $t<0$ and  $K=\L\w\kappa\in H^2\cap
H^\infty$.
\end{definition}

\par
We will  use the notation ``$\circ$''  for the convolution in $L_2(\R)$.

We are going to study linear predictors for anti-causal convolutions
$y=\kappa\circ x$ with $\kappa\in\K$. More precisely, we will study possibility
of  their approximation by causal convolutions  $\w y=\w\kappa \circ x$  with $\w\kappa\in\w\K$.
By the choice of $\K$ and $\w\K$, it follows that
 \baaa y(t)=\int_{t}^{+\infty}
\kappa(t-s)x(s)ds,\quad \w y(t)=\int_{-\infty}^t\w\kappa(t-s)x(s)ds.
\eaaa
The corresponding  predictors are linear; they are represented by causal time-invariant convolutions and  allow frequency representations  via transfer functions which is a preferable in electronic engineering, systems and control. This makes them convenient for applications.   In particular, this is because the linear time-invariant systems they can be realised via fixed  electronic hardware schemes.

\par
For $p\in[1,+\infty]$, we define linear normed spaces  $\Y_p$ of complex valued functions such that  $\Y_\infty\defi C(\R)$ and  $\Y_p\defi L_p(\R)$ for $p\in [1,+\infty)$.
\begin{definition}\label{def1} Let
$p\in[1,+\infty]$ be given.
Let  $\oo\X\subset \Y_p$ be a given set of functions $x:\R\to\R$.
\begin{itemize}
\item[(0)]
 We say that the set $\oo\X$ is  predictable at time $s\in\R$
 if, for any $x_1,x_2\in\oo\X$, if $x_1(t)=x_2(t)$ for a.e. $t<s$ then $x_1(t)=x_2(t)$ for a.e. $t\in\R$ .
\item[(ii)]
 We say
 that the set $\oo\X$ is  linearly $\Y_p$-predictable in
the weak sense  if, for any $\kappa\in\K$,  there exists a
sequence $\{\w\kappa_j\}_{j=1}^{+\infty}=\{\w\kappa_j(\cdot,\oo\X,\kappa)\}_{j=1}^{+\infty}\subset \w\K$ such that \baa\|y-\w
y_j\|_{\Y_p}\to 0\quad \hbox{as}\quad j\to+\infty\quad\forall
x\in\oo\X,\label{predict} \eaa where $y= \kappa\circ x$ and $\w y_j= \w\kappa_j \circ x$.
\item[(ii)] Let  $\oo\X$ be a set of processes which is also a
linear normed space  provided with a norm $\|\cdot\|_{\oo\X}$.
 We say that the set   $\oo\X$ is  linearly $\Y_p$-predictable in
the weak sense  uniformly with respect to the norm $\|\cdot\|_{\oo\X}$, if,
for any $\kappa\in\K$ and $\e>0$, there exists $\w\kappa =\w\kappa
(\cdot,\oo\X,\kappa,\|\cdot\|,\e)\in \w\K$ such that $$ \|y- \w
y\,\|_{\Y_p}\le \e\|x\|_{\oo\X}\quad \forall x\in\oo\X, $$ where
$y=\kappa\circ x$  and $\w y=\w\kappa\circ x$.
\end{itemize}
\end{definition}
\par
We call functions $\w\kappa_j$  and $\w\kappa$ in Definition \ref{def1}
predicting kernels.

\begin{proposition}\label{corrE}
Let $\oo\X$ be such as in Definition \ref{def1}(ii) with $p=2$. Then
 $\oo\X$  is predictable in the sense of  Definition \ref{def1}(i).
 \end{proposition}

The proof of Proposition \ref{corrE} given below  is based  on the completeness of the  system $\w\K$ in $L_2(-\infty,0)$.
In fact, even a smaller   set of finite linear combinations of exponents $e^{\lambda_k t}\Ind_{t<0}$, $k=1,...,\infty$ with $\lambda_k>0$ such that $\sum_{k} \lambda_k(1+\lambda_k^2)^{-1}=+\infty$
 is everywhere dense in  $L_2(-\infty,0)$; see e.g. \citet{Crum,Sed}. In theory, this may provide an approximate linear
 prediction method for the entire paths of processes being predictable in the sense of  Definition \ref{def1}(ii). For example,
 assume that the path $x|_{t\le 0}$ is observable. Then, for $t>0$,  a prediction $\w x(t)$ of $x(t)$
 can be approximated as  $\w x(t) \approx  \sum_k \w\xi_k f_k(t)$, where
 $\{f_k\}_{k=1}^\infty$ is an orthonormal basis in  $L_2(-\infty,0)$ constructed from the sequence $\{e^{\lambda_k t}\Ind_{t<0}\}_{k=1}^\infty$ by the Gram-Schmidt orthonormalization procedure, and where the values $\w \xi_k$
are the predictions of the integrals $\xi_k=\int_0^\infty x(s) f_k(- s)ds$ that can be found under the assumptions of  Definition \ref{def1}(ii).  This would be numerically challenging since  the predictors have to be constructed  for each $f_k$ individually.
In this paper, we focus  on the prediction  of single anti-causal convolutions.

The following examples  illustrate  the difference between different types of predictability
in Definition \ref{def1}.
\begin{example}\label{ex1}
\begin{enumerate}
\item
Any singleton set $\oo\X$  is predictable at any time in the sense of  Definition \ref{def1}(i).
\item Let $a>0$, and let  $x_0(t)=e^{-at}\Ind_{\{t\ge 0\}}$.
The singleton set $\{x_0\}$ is predictable at any time  in the sense of  Definition \ref{def1}(i) but
 is not  linearly predictable in the sense of  Definition \ref{def1}(ii) or  Definition \ref{def1}(iii).
\item Let $\O>0$ be given, and
let $\oo\X_\O$ be the set of all band-limited processes $x\in L_2(\R)$ such that
$X(i\o)=0$ if $|\o|>\O$, where $X=\F x$. Then  $\oo\X_\O$  is  linearly predictable in the sense of  Definition \ref{def1}(ii).
\item Let $\O>0$ be given, and
let $\ww\X_\O$ be the set of all high-frequency processes $x\in L_2(\R)$ such that
$X(i\o)=0$ if $|\o|<\O$, where $X=\F x$. Then  $\ww\X_\O$   is linearly predictable in the sense of  Definition \ref{def1}(ii).
\item
Let $\lambda>0$, and let  $x(t)=\Ind_{\{t\ge 0\}}e^{-\lambda t}$.
Let  a domain  $D_1\subset \R$ be given  such that
 $\mes D_1\in (0,+\infty)$. Let  $D_2\defi \R\setminus D_1$.
Let  $x_m=\F^{-1} X_m$, $m=1,2$, where
\baaa
X_m(i\o)=X(i\o)\Ind_{D_m}(i\o), \quad m=1,2,\quad X=\F x.
\eaaa
 Then the twin set $ \{x_1,x_2\}$ is predictable at any time  in the sense of  Definition \ref{def1}(i) but is not linearly predictable in the sense of Definition \ref{def1}(ii) with $p=2$.
\end{enumerate}
\end{example}
Example \ref{ex1}(v)  implies that  any larger class containing $\{x_1,x_2\}$  is
not linearly predictable in the sense of Definition \ref{def1}(ii).

It can be noted that processes from $\oo\X_\O$  with a interval spectrum gap at zero feature frequent oscillations (see \cite{BLU}), and yet Example \ref{ex1}(i) states that  this set is linearly predictable in the sense of  Definition \ref{def1}(ii).

\section{The main result}\label{secM}
For $q>0$, $c> 0$, and $\o\in\R$, set \baaa
h(\o,q,c)\defi \exp\frac{c}{|\o|^{q}}.
\label{hdef}\eaaa
\par
Let $\X(q,c)$ be the class of all processes $x\in L_2(\R)$ such that
\baa \|x\|_{\X(q,c)}\defi \esssup_{\o\in\R} |X\ew| h(\o,q,c)< +\infty, \quad \brea\hbox{where}\quad X=\F x.
\label{hfin}\eaa
This class includes processes from  $x\in L_2(\R)$ such that  their Fourier transforms vanish at $\o=0$
with the rate defined by $h$.

We consider $\X(q,c)$ as a linear normed space with the corresponding norm.
\par
 Note that $h(\o,q,c)\to +\infty$ as $\o\to 0$ and that (\ref{hfin}) holds for processes with spectrum degeneracy such that $X\ew$ is
approaching zero  as $\o\to 0$ with a sufficient rate of decay.
In particular, the class $\X$ includes  all band-limited processes
$x\in L_2(\R)$ such that there exists $\oo\O>0$ such that $X\ew=0$ for $\o\notin[-\oo\O,\oo\O]$, where $X=\F x$. However,
the spectrum degeneracy for functions from $\X$ is mild compared with the band-limitiness; in particular, these functions  are not necessarily  analytic, and their Fourier transform can be non-zero for all $\o\neq 0$.

\begin{example}\label{ex2}
\begin{enumerate}
\item  For any $q\ge 1$ and $c>0$,  the class $\X(q,c)$ is predictable in the sense of Definition \ref{def1}(i).
\item
If either $q\in(0,1)$ or $c\le 0$, then the class
$\X(q,c)$  is  not predictable  in the sense of Definition \ref{def1}(i).
\end{enumerate}
\end{example}

 Theorems \ref{Th1} and \ref{ThM} below   give, for the case where $q>1$,  a  constructive method of predicting of future averages of the processes descried via
convolutions;  for example, the values  $\int_0^\infty e^{-at}x(t)dt$ can be predicted for $a>0$ using the observations $x|_{t<0}$ and these predictors. Moreover, it is shown in Section \ref{secRob}  below that this prediction is robust with respect to the noise contamination.
These results represent extension of  the result \cite{D08} on the case of processes
 with a single point spectrum degeneracy.

 Let $\X\defi\cup_{q>1,c>0}\X(q,c)$,
\begin{theorem}\label{Th1} Let either $p=2$ or $p=+\infty$.
\begin{itemize}  \item[(i)] The class $\X$ is linearly $\Y_p$-predictable in the weak sense
such as described in Definition  \ref{def1}(ii).
\item[(ii)]  For any $c_0>0$ and $q_0>1$, the class $\X(q_0,c_0)$ is linearly
$\Y_p$-predictable  in the weak sense uniformly with respect to the
norm $\|\cdot\|_{\X(q_0,c_0)}$ such as described in Definition
\ref{def1}(iii).
\end{itemize}
\end{theorem}

\comm{
\begin{remark} It can be noted
that $x_m$ described  in Example \ref{ex1} cannot belong to $\X(q_0,c_0)$ simultaneously  for $m=1,2$.  \end{remark}
}

The predictability stated in Theorem \ref{Th1} is equivalent to the existence
of certain predicting kernels. The required kernels are presented  explicitly in the following theorem.
\begin{theorem}\label{ThM}
 Let $\kappa\in\K$ be given and represented as (\ref{kda}) for some given
$m>0$ and $\{a_j\}_{j=1}^m\subset (0,+\infty)$.  Let $r>2/(q-1)$ be
given. For $\g>0$ and $z\in\C^{+}\cup (i\R)$, set \baaa
&&V_j(z)\defi
1-\exp\left(-\g\frac{z-a_j}{z+\g^{-r}}\right),\quad\breakk
V(z)\defi \prod_{j=1}^mV_j(z),\\
  &&\w K(z)\defi V(z)K(z),\quad \w\kappa\defi \F^{-1}(\w K|_{ i\R}).
  \eaaa
 Then $\w K\in H^\infty\cap H^2$, and, for any sequence  $\g=\g_j\to +\infty$, the corresponding  sequence of kernels $\w\kappa$
ensures prediction required in Theorem \ref{Th1} (i)-(ii).
\end{theorem}

In particular, by the Paley-Wiener Theorem, it follows that $\w\k(t)=0$ for $t<0$, where $\w\k=\F^{-1} \w K(i\o)$.
Also, we  have that $\w\k=\F^{-1}\w K$ is real valued, since $\k$ is
real valued and
$K\left(-i\o\right)=\overline{K\left(i\o\right)}$, $V\left(-i\o\right)=\overline{V\left(i\o\right)}$.
\index{Note that $\g^{-r}\to 0$  as $\g\to+\infty$.}
\par Predicting kernels  $\w k$  in Theorem \ref{ThM} represent a modification
of the construction introduced in
\cite{D08} for continuous time processes with the spectrum vanishing on an interval.

\begin{remark} Since predicting kernels $\w\kappa$ in Theorem \ref{ThM} are real valued, it follows that
the corresponding processes $\w y=\w\kappa \circ x$  are real valued if $x$ is real valued.
This implies that Theorems \ref{Th1}-\ref{ThM} hold with a modification of Definition \ref{def1} involving real valued processes
$x,y,\w y_j$, and $\w y$.
\end{remark}

Any particular  predictor described in Theorem \ref{ThM} is not error-free and
ensures predictability in an approximate sense  only. However, the error  $\e$
can be done arbitrarily small; this can be achieved by  selection of a large enough $\g$.

The predictors in Theorem \ref{ThM}  do not depend on the polynomial $d$ in (\ref{kda}); however, they depend on $m$ and $\{a_k\}_{k=1}^m$ in (\ref{kda}).

The rate of spectrum vanishing for predictable processes considered in Theorem \ref{Th1}  is characterized by the pairs $(q,c)\in(1,+\infty)\times (0,+\infty)$.
The following proposition shows that the choice of
the critical values  here  is sharp.

\section{On robustness of the predictors  with respect to noise
contamination}\label{secRob} Let us show that the predictors introduced in Theorem \ref{ThM} and
designed for processes from $\X$ feature some robustness
with respect to noise contamination.   Suppose that these predictors  are
applied to a process $x\in L_2(\R)$ with a small noise
contamination
such that $x=x_0+\eta$, where $x_0\in\X$, and where
$\eta\in L_\infty(\R)\cap L_2(\R)$ represents the
noise. Let $X=\F x$, $X_0=\F x_0$, and $N=\F \eta$. We
assume that
$X_0(i\cdot)\in L_1(\R)$  and $\|N(i\cdot)\|_{L_1(\R)}=\nu$; we can write this as
 $X_0\in L_1(i\R)$ and
  and that $\|N\|_{L_1(i\R)}=\nu$. .  The parameter $\nu\ge 0$ represents the
intensity of the noise.
\par
By the assumptions,  the predictors  are constructed as in Theorem \ref{ThM}
under the hypothesis that $\nu=0$, i.e. that $\eta=0$ and
$x=x_0\in\X$. By Theorems \ref{Th1}-\ref{ThM}, for an arbitrarily small $\e>0$, there exists $\g$ such
that, if the hypothesis that $\nu=0$ is correct, then
\baaa \|\w
y-y\|_{L_{\infty}(\R)}\le\e,\label{eps}\eaaa
where $y$ and $\w y$ are such as in Definition \ref{def1}.
Let us estimate the prediction error for the case where $\nu>0$. We
have that \baaa \|\w y-y\|_{L_{\infty}(\R)}\le J_0+ J_{\NN},\eaaa
where\baaa &&J_0=\frac{1}{2\pi}\|(\w
K\ew-K\ew)X_0\ew\|_{L_1(\R)},\quad\breakk
J_{\NN}=\frac{1}{2\pi}\|(\w K\ew-K\ew)N\ew\|_{L_1(\R)}.
\eaaa The value $J_{\NN}$ represents the additional error caused by
the presence of unexpected high-frequency noise (when $\nu>0$). It
follows that \baa \|\w y-y\|_{L_{\infty}(\R)}\le
\e+\nu(\varkappa+1),\label{yn}\eaa
where $\varkappa\defi \sup_{\o\in\R}|\w K\ew|$.
\par
Therefore, it can be concluded that the prediction  is robust with
respect to noise contamination for any given $\e$. On the other
hand, if $\e\to 0$ then $\g\to +\infty$ and $\varkappa\to +\infty$. In
this case, error (\ref{yn}) is increasing for any given $\nu>0$.
Therefore, the error in the presence of noise will be large for a predictor targeting too small a size of the
error for the noiseless  processes from $\X$.
\par
The equations describing the dependence of $(\e,\varkappa)$ on $\g$
could be derived similarly to estimates in \cite{D12}, Section 6,
where it was done for different predicting kernels and for
band-limited processes. We leave it for future research.

\section{Proofs}\label{secProof}
{\it Proof of Proposition \ref{corrE}}. Let us prove statement (i) first.  It suffices to show that if $s\in\R$ and $x\in\oo\X$ are such that $x|_{t\le s}=0$, then
$x|_{t> s}=0$.

Suppose that there exists $s\in\R$ and $x\in\oo\X$ such that $x|_{t\le s}=0$.
For $\kappa\in\K$, let  $y$, $\w y_j$, and $\w\kappa_j$, be such as described  in Definition \ref{def1} (i).
Since  $x|_{t\le s}=0$, it follows that $\w y_j(s)=0$ for any $j$ and any $s<0$. On the other hand, (\ref{predict}) holds by the assumption on $\oo\X$  in  statement (i).
Hence $y(s)=0$ for any $\kappa\in\K$.
Furthermore,  the class $\K$ contains functions $\kappa(t)=e^{\lambda t}\Ind_{t\le 0}$ for all $\lambda>0$; it follows for these functions  that
\baaa
y(s)=\int_s^{\infty}e^{\lambda(s-t)}x(t)dt=\int_s^{\infty}e^{-\lambda(t-s)}x(t)dt =0 \quad \brea\forall \lambda>0.
\eaaa
 The M\"untz-Sz\'asz Theorem implies that there exits a set $\{\lambda_k\}_{k=1}^\infty\subset (0,+\infty)$ such that
  that  the  set of finite linear combinations of exponents $e^{-\lambda_k t}$ is complete in $L_2(s,+\infty)$, meaning that the
  set of finite linear combinations   of these exponents is everywhere dense in  $L_2(s,+\infty)$; see e.g. \citet{Crum,Sed}.
It follows that
$x|_{t> s}= 0$.
This completes the proof of Proposition \ref{corrE}(i).

{\em Proof for Example \ref{ex1}.}
The proof for  Examples \ref{ex1}(i-ii) is obvious.  The proof for  Examples \ref{ex1}(ii-iv) is given in \cite{D08}.

Let us prove Example \ref{ex1}(v).
We have that $x_1(t)=-x_2(t)$ for  a.e. $t\le 0$, and that the process $x_1(t)$ is band-limited and hence  continuous.

Suppose  that  $x_1(t)=x_2(t)$ for  a.e. $t\le s$ for some $s\in\R$.  It would imply that  $x_1(t)=x_2(t)=0$ for  $t\le s$.
Thi is impossible since since  $x_1\neq 0$ and $x_1$ is a band-limited process, it follows that $x_1$ cannot vanish on an open interval; otherwise, it its unique analytic extension would be zero.
Therefore, we have  proved   that the set $\{x_1,x_2\}$ is   predictable at any time  in the sense of  Definition \ref{def1}(i),

Let us show  that the set $\{x_1,x_2\}$ is   not predictable at any time  in the sense of  Definition \ref{def1}(ii).

Let $\kappa\in\K$ be fixed, and let $y_m=\kappa\circ x_m$, $m=1,2$.
Suppose that there exist kernels $\w\kappa_j\in\w\K$ required in Definition \ref{def1}(i) for $\V$.
Let  $\w y_{m,j}=\w\kappa_j\circ x_m$ and $\w Y_{m,j}=\F\w y_{m,j}$.

   We have that
  \baaa
  2\pi\|y_m-\w y_{m,j}\|_{L_2(R)}^2\breakk
  =\int_{\R}\Ind_{D_{m}}(\o)|K(i\o)-\w K_{j}(i\o)|^2|X_m(i\o)|^2d\o
  \eaaa
for $m=1,2$. Hence
 \baa
 &&2\pi\sum_{m=1,2}\|y_m-\w y_{m,j}\|_{L_2(R)}^2\breakk
 =2\pi\sum_{m=1,2}\int_{\R}|K(i\o)-\w K_{j}(i\o)|\Ind_{D_{m}}(\o)|X_m(i\o)|^2 d\o
\nonumber \\
  &&
  =2\pi\int_{\R}|K(i\o)-\w K_{j}(i\o)(\o)|X(i\o)|^2 d\o\nonumber
\\
 && =\int_{\R}[K(i\o) -\w K_{j}(i\o)]X(i\o)\overline{[K(i\o) -\w K_{j}(i\o)] X(i\o)}d\o. \label{sum}
\eaa

Let  $C^-\defi \{z\in\C: \ \Re z<0\}$, and let  $H^2_-$ be the set of functions $F(z)$ defined on $\C^-$
such that $F(-\oo z)\in H^2$; the inverse  Fourier transforms $f\in L_2(\R)$ of these  functions are
 such that $f(t)=0$ for $t>0$.

By (\ref{sum})-(\ref{QQ}), it follows that
 \baaa
  &&2\pi\sum_{m=1,2}\|y_m-\w y_{m,j}\|_{L_2(R)}^2\breakk
  = \left((K(i\o)-\w K_{j}(i\o)X(i\o),(K(i\o)-\w K_{j}(i\o))X(i\o)\right)_{L_2(\R)}
\\  &&
   = \|K(i\o)X(i\o)\|_{L_2(\R)}^2
   +\|\w K_{j}(i\o))X(i\o)\|_{L_2(\R)}^2-2R,
\eaaa
where
\baaa
&&R
=\Re \int_{\R}\overline{K(i\o)}\overline{ X(i\o)}  X(i\o)\w K_j(i\o) d\o
\breakk=\Re\int_{\R}\overline{Y(i\o)}   \w K_j(i\o)X(i\o)d\o,
\eaaa
where $Y(z)\defi K(z) X(z)$.

  Assume that $K\in\K$ be such that $K(z)=d(z)/\d(z)$
 where $\d(z)$ is a polynomial of order $m>1$ with the roots containing in the set $\{z\in \C:\ \Re z>0,\ z\neq \lambda\}$,
 and where
$d(z)= (z+\lambda)d_0(z)$ for  a non-zero polynomial $d_0(z)$ such that $\deg d_0< m-1$.
By the choice of $x$, we have that   $X(i\o)=(\lambda+i\o)^{-1}$; this implies that  $Y=KX\in H^2_-$.
By the orthogonality in $L_2(i\R)$ of the traces of functions  from Hardy spaces  $H^2$ and $H_-^2$ respectively, we obtain that
\baa
&&R=\Re \left(Y(i\o),\w K_{j}(i\o) X(i\o)\right)_{L_2(\R)} \breakk =\Re \left(K(i\o)X(i\o),\w K_{j}(i\o) X(i\o)\right)_{L_2(\R)}
=0.
\label{QQ}\eaa
By (\ref{sum})-(\ref{QQ}), it follows that
 \baa
  &&2\pi\sum_{m=1,2}\|y_m-\w y_{m,j}\|_{L_2(R)}^2
  \\&&=\int_{\R}(|K(i\o) |^2 -2 \oo K(i\o)  {\w K}(i\o) +|\w K_j(i\o) |^2 )|X(i\o)|^2d\o . \hphantom{xx}\label{yy}\eaa
It follows from (\ref{yy}) that  any choice of $\w\kappa_j$ cannot  ensure that $\|y_m-\w y_{m,j}\|_{L_2(\R)}\to 0$
simultaneously  for $m=1$ and $m=2$, which is inconsistent with the  supposition that  conditions in Definition \ref{def1} are satisfied for the set
 $\{x_1,x_2\}$.
This completes the proof of  Example \ref{ex1}.
 $\Box$

 It can be noted that both singletons $\{x_1\}$ and  $\{x_2\}$ defined in Example \ref{ex1}(v)
are  linearly predictable in the sense of Definition \ref{def1}(ii) with $p=2$, and yet the twin set $\{x_1,x_2\}$
is not linearly predictable in this sense.

\par
{\em Proof of Example \ref{ex2}}.
  It is known that if $x\in L_2(\R)$ and $x|_{t<0}=0$ a.e. then $X=\L x\in H^2$ and $\int_{-\infty}^\infty \log |X(i\o)|(1+\o^2)^{-1}d\o>-\infty$; see, e.g. Theorems  11.6 and 11.7 from \cite{Du}.
This implies that,  for any  $q\ge 1$ and $c>0$, $\X(q,c)\cap H^2=\{0\}$. Hence it  cannot happen simultainuously
that $x=x_1-x_2\neq 0$,  $x_1,x_2\in\X(q,c)$, and $x\in H^2$ (i.e. $x(t)=0$ for $t<0$) .
This   implies that the class $\X(q,c)$ is predictable in the sense of Definition \ref{def1}(i).

Let us prove Example \ref{ex2}(ii).
For any $c\le 0$, by the definitions, $\X(q,c)$ is the class of $x\in L_2(\R)$ such that $X=\F x\in L_\infty(i\R)$; obviously, this class  is too wide and cannot be predictable
in the sense of Definition \ref{def1}(i-iii).
Therefore, it suffices to consider $q\in(0,1)$ and $c>0$ only.

Assume that $q\in (0,1)$ and $c>0$ be given.
Consider a filter with the transfer function $\Psi(z)\defi (1+z)^{-2}\ww\Psi(z)$, where  $\ww\Psi(z)\in H^\infty$
is such that
$|\ww\Psi\ew|=\exp(-c|\o|^{-q})$, $\o\in\R$. Since $q<1$, we have that $(1+\o^2)^{-1}\log|\ww\Psi\ew|\in L_1(\R)$. Hence
such $\ww\Psi$ exists; see,  e.g.  Theorem 11.6 in \cite{Du}, p. 193. By the choice of $\Psi$,  this
filter is causal.
Let \baaa
\X_\psi\defi \{x\in L_2(\R):\   X\ew=\Psi\ew \ww Y\ew,\ \brea X=\F x,\  \ww Y=\F \ww y,\ \ww y\in \X(q,0)\}.
\eaaa

Suppose that the class $\X(q,c)$ is  linearly predictable  in the sense of Definition \ref{def1}(ii).
By the definitions,  $\X_\psi\subseteq\X(q,c)$, hence  the class $\X_\psi$ should be also linearly predictable  in the sense of Definition \ref{def1}(ii).
On the other hand, $\X_\psi$ consists of processes from $\X(q,0)$ transformed
by a causal filter.
As was mentioned above, the class $\X(q,0)$  cannot be linearly  predictable.
 Therefore,  the class $\X_\psi$ also is not linearly predictable in the sense of Definition \ref{def1}. Hence the supposition is incorrect and the
 class $\X(q,c)$ cannot be linearly predictable in this sense for $q\in (0,1)$.
This completes the proof of Example  \ref{ex2}. $\Box$.

\vspace{4mm}
To proceed further, we need  to establish some properties of the function $V$.
\par
 Let $\kappa\in \K$ and the corresponding set $\{a_k\}_{k=1}^m\subset (0,+\infty)$ be given.
 Let $\oo a\defi \max_{j=1,...,m}a_j$, $\O(\g)=\sqrt{\oo a\g^{-r}}$, let $D(\g)=[-\O(\g),\O(\g)]$, and let
 $D_+(\g)\defi\R\backslash D(\g)$.
\begin{lemma}\label{lemmaV}
\begin{itemize}
\item[(i)] $V\in H^{\infty}$ and $\w \Ko \defi \Ko V\in
H^{\infty}\cap H^2$ for any $\g>0$.
\item[(ii)]  $\Re \left(\frac{i\o-a_j}{i\o+\a}\right) >0$ and $|V_j\ew -1|<1$ for any $\g>0$, $\o\in D_+(\g)$, and $j\in\{1,...,m\}$.
\item[(iii)] $V(i\o)\to 1$  as  $\g\to +\infty$ for all  $\o\in \R\setminus\{0\}$.
\item[(iv)] For any $q>1$ and $c>0$,
there exists $\g_0>0$ such that  $|V\ew|
h(\o,q,c)^{-1}\le 1$ for any $\g\ge \g_0$ and $\o\in D(\g)$.
\end{itemize}
\end{lemma}
\par

{\it Proof of Lemma \ref{lemmaV}}.  Clearly,
\baaa 1-\exp(z)=-\sum_{k=1}^{+\infty}\frac{(-1)^k z^k}{k!}.  \eaaa
Hence  \baaa
V_j(z) =-\sum_{k=1}^{+\infty}\frac{(-1)^k\g^k(z-a_j)^k}{k!(z+\g^{-r})^k} \brea=-(z-a_j)\sum_{k=1}^{+\infty}\frac{(-1)^k\g^k(z-a_j)^{k-1}}{k!(z+\g^{-r})^k}.
\eaaa
Hence $V\in
H^{\infty}$. It also follows  that $\d(z)^{-1}V(z)\in H^2\cap H^{\infty}$, since
each pole at $z=a_k$ of $\d(z)^{-1}$  is being compensated by multiplying on $V_j(z)$.  Then
statement (i) follows from the Paley-Wiener
theorem.
\par Further, we have for $\o\in\R$ and $j=1,...,m$ that
 \baaa
&&\frac{i\o-a_j}{i\o+\g^{-r}}=\frac{(-a_j+i\o)(\g^{-r}-i\o)}{\o^2+\g^{-2r}}\breakk=\frac{\o^2-a_j\g^{-r}}{\o^2+\g^{-2r}}
+i\frac{-a_j \o+\g^{-r}\o}{\o^2+\g^{-2r}}\,. \eaaa
 Hence
 \baa
\Re\frac{i\o-a_j}{i\o+\g^{-r}}
=\frac{\o^2-a_j\g^{-r}}{\o^2+\g^{-2r}}\ge \frac{\o^2-\O(\g)^2}{\o^2+\g^{-2r}}>0, \quad \brea\o\in D_+(\g).
\label{re} \eaa
By the definitions, it follows  that
\baa
&&|V_j\ew-1|=\left|\exp\left(-\g\frac{i\o-a_j}{i\o+\g^{-r}}\right)\right|\breakk=\exp\left(-\g\Re\frac{i\o-a_j}{i\o+\g^{-r}}\right).
\label{V1<1}\eaa
Hence
\baaa
|V_j\ew-1|< 1,\quad \o\in D_+(\g).
\eaaa
This implies statement (ii).

Further,  $\g^{-r}\to 0$  and $\O(\g)\to 0$  as $\g\to+\infty$.
Hence, by (\ref{re}), there exists $M>0$ such that, for any $\nu>0$, there exists $\g_{\nu,M}>0$ such that
\baaa
\Re\frac{i\o-a_j}{i\o+\g^{-r}} \ge M, \quad \o\in\R\setminus(-\nu,\nu),\quad\brea \g\ge \g_{\nu,M},\quad j=1,...,m.
\label{re2} \eaaa
This and (\ref{V1<1}) imply statement (iii).
\par
Let us prove statement (iv). Let $\oo\g>0$ be selected, and let $\G\defi \sup_{\g\ge \oo\g,\ j=1,...,m}\sqrt{\O(\g)^2+a_j^2}$. For $j\in\{1,...,m\}$
and  $\o\in D(\g)$, we have
that
\baaa
\left|\Re\frac{i\o-a_j}{i\o+\g^{-r}}\right|\le \left|\frac{i\o-a_j}{i\o+\g^{-r}}\right|\le \frac{\G}{|\o|}
\eaaa
and \baaa|V_j(i\o)|h(\o,q,c)^{-1}\le
\exp\left(\g\left|\Re\frac{i\o-a_j}{i\o+\g^{-r}}\right|-
\frac{c}{|\o|^q}\right)\brea\le
\exp\left(\frac{\G\g}{|\o|}- \frac{c}{|\o|^q}\right)=\exp\frac{\G \g|\o|^{q-1}- c}{|\o|^q}
\\ \le\exp\frac{\G \g\O(\g)^{q-1}- c}{|\o|^q} \le\exp\frac{\G \g\cdot \oo a^{1/2} \g^{-r(q-1)/2}- c}{|\o|^q}=
\exp\frac{\G \oo a^{1/2} \g^{1-r(q-1)/2}- c}{|\o|^q}.
\eaaa
 By the assumptions on $r$, we have that  $1-r(q-1)/2<0$.
Hence,
for any $q>1$ and $c>0$, there exists $\g_0>0$ such that for any $\g\ge \g_0$
\baaa|V_j(i\o)|h\left(\o,q,\frac{c}{m}\right)^{-1} \le 1,\quad \o\in D(\g),\quad j=1,...,m. \eaaa By the choice of $h$, it follows that
\baaa h\left(\o,q,\frac{c}{m}\right)^m=h(\o,q,c).\eaaa
Hence \baaa  |V(i\o)|
h\left(\o,q,c\right)^{-1}= \prod_{j=1}^m |V_j(i\o)|
h\left(\o,q,\frac{c}{m}\right)^{-1} \le 1,\quad\brea \o\in D(\g).\eaaa
 This completes the proof of   statement (iv) and Lemma \ref{lemmaV}. $\Box$
\vspace{0.5cm}

{\it Proof of Theorem \ref{Th1}}. Theorem \ref{Th1}  follows immediately from Theorem \ref{ThM} which proof is given below. $\Box$

\par
{\it Proof of Theorem \ref{ThM}}. Let $\kappa\in \K$ be given, $K=\F \kappa$. Let $\g=\g_j\to +\infty$,  and let $(V,\w K)$  be the corresponding functions.

 Let $\ko=\F^{-1}\Ko $ and $\w
\ko=\F^{-1}\w \Ko$. For  $x\in \X$, let $X\defi \F x$ and
\baaa \yo (t)\defi \int_{t}^{\infty}\ko (t-s)x(s)ds,\quad \w
\yo (t)\defi \int^t_{-\infty}\w \ko (t-s)x(s)ds. \eaaa

Let
 $\Yo \ew\defi (\F \yo )\ew=K\ew X\ew$.
 By
the definitions, it follows that
 $\w Y\ew=\w K\ew X\ew$.
\par
Further, let  $\rho=2$ if $p=2$ and $\rho=1$ if $p=+\infty$.
\par
We have that $\|\w Y\ew-Y\ew\|_{L_\rho (\R)}^\rho=I_1+I_2,$
where \baaa &&I_1=\int_{D(\g)}|\w Y\ew-Y\ew|^\rho d\o,\qquad\breakk
I_2=\int_{D_+(\g)}|\w Y\ew-Y\ew|^\rho d\o. \eaaa

By the assumptions, there exists $c>0$ such that $\|X\ew
h(\o,q,c)\|_{L_\infty(\R)}<+\infty$. Hence
 \baaa I_1^{1/\rho}&=&\|\w
 Y\ew-Y\ew\|_{L_\rho (D(\g))}\breakk =
\|(\w K\ew-K\ew)X\|_{L_\rho (D(\g))}\nonumber\\
&\le& \|(V \ew -1)
 h(\o,q,c)^{-1}\|_{L_\rho (D(\g))}\breakkk \|K\ew X\ew h(\o,q,c)\|_{L_\infty(\R)}\nonumber
\\
&\le& \Bigl(\|V \ew
 h(\o,q,c)^{-1}\|_{L_\rho (D(\g))}+c_h(\g)\Bigr)\breakkk \|K\ew X\ew h(\o,q,c)\|_{L_\infty(\R)},
 \nonumber
 \eaaa
 where \baaa
 c_h(\g)\defi \|
 h(\o,q,c)^{-1}\|_{L_\rho (D(\g))}.
 \eaaa
  Clearly, $c_h(\g)\to 0$ as $\g\to +\infty$.
 Further, the measure of the set $D(\g)$ is
$2\sqrt{\oo a\a }$. By  Lemma \ref{lemmaV}
(iv), \baaa
\int_{D(\g)}|V_j\ew|^\rho
h(\o,q,c)^{-\rho}d\o\le 2\sqrt{\oo a\g^{-r}}\to 0\eaaa
as $\g\to 0$ for any $\rho\ge 1$ and any $j\in\{1,...,m\}$.
 It follows that \baaa
&&I_1^{1/\rho}\le \Bigl[\Bigl(2\sqrt{\oo a\g^{-r}}\Bigr)^{1/\rho}+c_h(\g)\Bigr]\breakkk\|K\ew \|_{L_\infty(\R)} \|X\ew h(\o,q,c)\|_{L_\infty(\R)}\to 0\quad\breakk \hbox{as}\quad \g\to +\infty.
\label{4s}\eaaa
Therefore, $I_1\to 0$ as $\g\to +\infty$.
\par
Let us estimate $I_2$.
  We
 have that
\baaa I_2=\int_{D_+(\g)} |K\ew(1-V\ew) X\ew| ^\rho d\o\brea\le
\psi(\g)\|X\ew\|_{L_\infty(\R)}\\\le \psi(\g)\|X\ew
h(\o,q,c)\|_{L_\infty(\R)} ^\rho ,\eaaa where \baaa &&\psi(\g)
\defi\int_{D_+(\g)} |K\ew(1-V\ew)| ^\rho d\o\breakk=\int_{-\infty}^\infty
\Ind_{D_+(\g)}(\o) |K\ew(1-V\ew)| ^\rho d\o. \eaaa Here $\Ind$
denotes the indicator function.
\par
  By Lemma \ref{lemmaV}(iii),
$\Ind_{D_+(\g)}(\o) |K\ew(1-V\ew)| ^\rho \to 0$ a.e. as $\g\to
+\infty$. By Lemma \ref{lemmaV}(ii),
$|V_j\left(e^{i\o}\right)-1|\le 1$ for all $\o\in D_+(\g)$. Hence \baaa
\Ind_{D_+(\g)}(\o)|K\ew(1-V\ew) | ^\rho\brea \le 2^{m\rho} \sup_{\o\in
D_+(\g)}|K\ew | ^\rho .\eaaa   From Lebesgue Dominance Theorem,
it follows that $\psi(\g)\to 0$ as $\g\to+\infty$. It follows that
$I_1+I_2\to 0$ for any $q>1$ and $c>0$, $x\in\X(q,c)$. By the definition of
$\rho$, we have that $1/\rho+1/p=1$. Hence $\|\w y-y\|_{L_p(\R)}\to
0$ as $\g\to +\infty$ for any $x\in\X$.
  It follows that the
predicting kernels $\w\kappa=\F^{-1}\w K$ are such as required in statement (i) of Theorem \ref{Th1}.
This
completes the proof of statement (i).

Let us show that these kernels are such as required in statement (ii) of Theorem \ref{Th1}.  Let
\baaa
\xi(\g)\defi \Bigl(2\sqrt{\oo a\g^{-r}}\Bigr)^{1/\rho}+c_h(\g)+ \psi(\g).
\eaaa
 We have that \baaa \|\w
Y\ew-Y\ew\|_{L_\rho (\R)} ^\rho = I_1+I_2\brea\le \xi(\g)\|X\ew h(\o,q,c_0)\| ^\rho
_{L_\infty(\R)}\eaaa for any
$x\in\X(q_0,c_0)$.  It follows from the proofs above that $\xi(\g)\to 0$ as $\g\to +\infty$. Hence (\ref{predict}) holds
for the corresponding $y=\F^{-1}Y$ and $\w y_j=\F^{-1} \w Y_j$.   In addition,  it follows that the
predicting kernels $\w\kappa=\F^{-1}\w K$ are such as required in statement (ii) of Theorem \ref{Th1}.

Since $X\ew \in L_2(\R)\cap L_\infty(\R)$, $K\ew \in L_2(\R)\cap L_\infty(\R)$ and $\w K\in H^\infty \cap H^2$, it follows that $y\in C(\R)$ and $\w y\in  C(\R)$. For this $y $ and $\w y$, the
norms
in $L_\infty(\R)$ are the same as  the norms in $\Y_\infty=C(\R)$.
This completes the proof of Theorem \ref{ThM}. $\Box$

\section{Discussion and future research}\label{secCon}
The present paper is focused on the
impact of spectrum degeneracy at a single point for continuous time processes in pathwise deterministic setting. The  paper suggests  frequency  criteria  of a  linear predictability of anti-causal  convolutions and linear predictors
described explicitly in the frequency domain. The predictability is feasible for classes of processes with a single point spectrum degeneracity.
 \begin{enumerate}
\comm{\item
The suggested condition of linear predictability  reminds  the classical Kolmogorov-Krein  criterion (\ref{KK})  for the spectral densities  and
stochastic Gaussian processes. The presented result is not a straightforward rewording of this criterion. The paths
of these continuous time stationary Gaussian processes have very special features, and the spectral density
cannot be automatically associated with the Fourier transform.
}
 \item The  family of  predictors suggested in Theorem \ref{ThM}
 do not depend on the shape of the spectrum of the underlying process.  This
  could  be useful for applications.

\item
The predictors from Theorem \ref{ThM}  are not error-free; however, the error can be made
arbitrarily small with a choice of large $\g$. In addition, these  predictors  feature
robustness with respect to noise contamination.
If the predictor is targeting too small a size of
the error, the norm of the
transfer function  will be large; this could lead to a larger error caused by the presence of noise.
\item
There is some similarity with a result obtained in \cite{D12} for discrete time processes (sequences):
 they  are predictable if their Z-transforms vanish at a point of the unit circle $\T=\{z\in \C:\ |z|=1\}$. However, the result \cite{D12} was less unexpected  since a sequence
is band-limited and predictable  if its Z-transform vanishes on any arbitrarily small arc on $\T$.
\comm{ \item
It is yet unclear if the class $\X(1,c)$ features some predictability with $c>0$.
 We leave this  for the future research.}
\xxxonly{\item
For the processes  with infinite discrete  periodic sets of points of degeneracy,
the results of  this paper can be extended on prediction in the "strong" sense such that   $y(t)=x(t+\tau )$ in Definition \ref{def1}. Here  $\tau >0$ is a prediction horizon.
For this, statements similar to Theorems \ref{Th1}-\ref{ThM} and Corollary \ref{corrE}  hold  with the following replacements:
we have to select
 \baaa &&h(\o,q,c)=\exp\frac{c}{|e^{i \tau \o}+1|^{q}}, \qquad K(z)=e^{\tau p}, \quad p\in \C^+,\\
&&V(z)\defi 1-\exp\left(-\frac{\g}{e^{\tau p}+1-\g^{-r}}\right),\break\quad\w
K(z)\defi K(z)V(z). \eaaa
Similar transfer function was introduced in a discrete time setting in \cite{D12}; some numerical examples can be found in \cite{D16}.\index{ (in the notation of \cite{D12}, $r$ is replaced by $2\mu/(q-1)$).} Larger prediction horizon  $\tau $ would require
more frequent points of spectrum degeneracy $\{ (\pi+2\pi k)/ \tau \}_{k\in\ZZ}$ for underlying processes . We leave more detailed analysis for the future research.}
\item It is still unclear if the linear predictability is feasible for the class  $\X(1,c)$ with some $c>0$.
\item
The  processes with a interval spectrum gap at zero feature frequent oscillations (sign changes)  \cite{BLU};
it would be interesting to see if the processes from $\X$ have some similar properties.

\end{enumerate}

\noxxx{ \subsubsection*{Acknowledgment} This work was financially supported by Government of Russian Federation (Grant 08-08).}

\end{document}